\documentclass[12pt]{article}
\usepackage{hyperref}
\usepackage{amsmath}
\usepackage[utf8]{inputenc}  
\usepackage{amsmath}
\usepackage{enumitem}        
\usepackage{natbib}
\usepackage{mathtools}
\usepackage{amsmath}
\usepackage{amsfonts}
\usepackage{amssymb}
\usepackage{amsmath}
\usepackage{graphicx}
\usepackage{natbib}
\title{Statistical Analysis of Magnetic Patches in Solar Active Regions Using the Complex Network Method}

\author{
	Zahra Tajik\thanks{Department of Physics, Faculty of Science, University of Zanjan, University Blvd., Zanjan, 45371-38791, Iran. \texttt{Z.tajik@znu.ac.ir}} \and
	Hasan Rezaei\thanks{Department of Geography, Faculty of Basic Sciences, Imam Ali Nazaja University, Postal Code: 1317893471, Tehran, Iran} \and
	Mehdi Simiari\thanks{Department of Physics, Faculty of Science, Imam Ali University, Postal Code: 1317893471, Tehran, Iran} \and
	Hossien Safari\thanks{Department of Physics, Faculty of Science, University of Zanjan, University Blvd., Postal Code 45371-38791, Zanjan, Iran. Email: safari@znu.ac.ir}
}
\begin{document}

\maketitle 

\begin{abstract}
Identifying solar active regions (ARs), which consist of one or more pairs of magnetic patches with opposite polarities, is essential due to their significant role in dynamic solar atmospheric phenomena. In this study, we analyze ARs during their emergence and evolution on the solar surface using a complex network-based method known as Identifying Solar Magnetic Patches (ISMP). To examine the magnetic characteristics, we selected a subregion of 125 $\times$ 125 pixels centered on AR NOAA No., 1158, observed in 2011. Line-of-sight magnetogram data were obtained from the Helioseismic and Magnetic Imager (HMI) onboard the Solar Dynamics Observatory (SDO).
Our statistical analysis reveals that the distributions of patch area, lifetime, and magnetic flux follow power-law behavior, with exponents approximately equal to $\alpha$ = 2.14, 2.5, and 1.42, respectively. Furthermore, a Hurst exponent of 0.57 indicates the presence of long-range temporal correlations in the emergence of new magnetic patches.
\end{abstract}


\section{Introduction}
The definition of solar ARs has evolved significantly with advancements in observational tools and multi-wavelength imaging techniques. Currently, an AR is understood as a three-dimensional volumetric region where the solar magnetic field evolves from the photosphere to the corona — the primary location for most solar magnetic activity \citep{van}. Magnetic phenomena such as solar flares and coronal mass ejections (CMEs) are widely considered direct outcomes of topological and structural reconfigurations within the magnetic fi eld of ARs \citep{sammis, periest, Aschwanden, falconer,  mason2010, georolis,  abramenko, barnes, raboonik, Farhad}. Therefore, identifying ARs and investigating their statistical properties and magnetic evolution can provide crucial insights into the mechanisms driving these explosive events.

To date, extensive efforts have been made to characterize solar magnetic patches as the fundamental building blocks of solar magnetic fields. These investigations have largely focused on statistical properties such as magnetic flux, area, and lifetime \citep[e.g.,][]{hagenaar2003, abramenko2005,Gosic2016}. \citet{Schrijver1997a} introduced the concept of the ``magnetic carpet,'' highlighting the continual emergence, migration, and cancellation of small-scale magnetic features across the solar surface. Subsequent studies revealed that small-scale magnetic flux exhibits power-law distributions and reflects the evolving nature of solar activity throughout the solar cycle \citep{meunier2003statistical}. 

\citet{hagenaar2003} reported that AR magnetic elements typically possess fluxes in the range of $10^{18}$ to $10^{20}$~Mx and follow a log-normal distribution. \citet{abramenko2005} demonstrated that magnetic regions with flatter spectral slopes tend to show greater flare productivity, while \citet{wang2008} identified a relationship between high-flux, topologically complex structures and the occurrence of fast CMEs. A comprehensive study by \citet{parnell2009power} found that magnetic flux concentrations across different scales follow a scale-invariant power-law distribution with a slope of approximately $-1.85$, suggesting underlying self-similar mechanisms.

 \citet{javaherian2017} applied YAFTA methods to over 185,000 quiet-Sun magnetic features in SDO/HMI data and showed that approximately 95\% of these patches vanish within 100 minutes. 
 
 In recent years, machine learning algorithms have also been employed to identify and investigate ARs and predict their flare potential based on extracted magnetic parameters \citep{zhang2022, chen2024}. While previous studies have extensively characterized the static properties of magnetic patches, their dynamic evolution remains poorly understood. Using a complex network method, this study statistically investigates the properties of magnetic patches in solar ARs, including their distributions of area, flux, and lifetime, as well as their spatial organization and topological characteristics.

To investigate magnetic patches across multiple spatial and temporal scales, we applied a complex network-based (ISMP algorithm) to segment, track, and extract the physical parameters (e.g., size distribution, filling factors, and magnetic flux) of both negative and positive polarities from HMI/SDO magnetograms taken in 2011. These temporal networks enable a statistical examination of the evolving magnetic connectivity throughout the lifetime of the AR. As a case study, we analyze a 570.75 minute time series of AR NOAA No., 1158 observed during its disk-center passage on 14–15 February 2011.

The structure of the paper is as follows: Section~\ref{sec:data} describes the dataset; Section~\ref{sec:method} details the identification and tracking technique; Section~\ref{result} presents the results along with their interpretation; and Section~\ref{sec:conclusion} provides the concluding remarks.

\section{Data and Observations}\label{sec:data}
Understanding the dynamic behavior of solar magnetic fields demands continuous and high-resolution observations, particularly of the photosphere where these fields are generated and evolve. SDO, a space-based mission dedicated to observing the Sun, has played a transformative role in this regard. Among its instruments, HMI is especially important for capturing detailed, uninterrupted measurements of the solar photosphere \citep{scherrer, schou2012design}. 

The HMI instrument is designed to study the solar magnetic field and its temporal evolution with exceptional precision. It provides full-disk observations at three different spatial resolutions—$4096 \times 4096$, $2048 \times 2048$, and $1024 \times 1024$ pixels—by imaging the Sun in the $Fe\,\textsc{i}$ absorption line at $6173$~\AA. These observations are made with a spatial resolution of approximately $1^{\prime\prime}$ and a typical cadence of 45 and 75 seconds, depending on the data product \citep{Pesnell, DeRosa}. HMI provides a variety of observational products, including continuum filtergrams, Dopplergrams, and both line-of-sight (LOS) and vector magnetograms, which collectively enable a comprehensive study of solar surface phenomena. All data are publicly accessible via the Joint Science Operations Center (JSOC) at \url{http://jsoc.stanford.edu}.\\
To place our study in context, the temporal and spatial evolution of ARs on the solar surface is also documented by international solar monitoring services, such as the Royal Observatory of Belgium (ROB), the Solar Influences Data Analysis Center (SIDC), and the National Oceanic and Atmospheric Administration (NOAA). Each AR is assigned a unique four-digit NOAA number—a system that has been in use since 1972 \citep{aurora}.\\
In this study, we utilize the HMI/LOS magnetograms with a spatial sampling of $2.4^{\prime\prime}$ pixels$^{-1}$ at $1024 \times 1024$ pixels. The dataset consists of a continuous 570.75 minutes (approximately 9.5 hours) time series with a cadence of 45 seconds, spanning from 17:00:45 UTC on 14 February 2011 to 02:30:45 UTC on 15 February 2011. To minimize projection effects, we extracted a $125 \times 125$ pixel subregion centered on AR NOAA No., 1158, located near the center of the solar disk during the observation period. AR NOAA No., 1158 was selected due to its intense magnetic activity, including major X-class flares, making it an ideal case for analyzing the dynamic evolution of magnetic patches \citep{Pesnell2011}.

\section{Method}\label{sec:method}
Given that solar ARs are the source of phenomena in the solar atmosphere, identifying and studying their magnetic evolution is crucial and has garnered significant research interest. Investigating these regions enhances our understanding of solar magnetic phenomena and could potentially aid in forecasting. In this study, we implement the process of identifying and tracking magnetic elements on the solar surface using the method presented in \cite{tajik2024complex}. This method constructs a complex network from the magnetogram data of the HMI instrument onboard the SDO spacecraft. We then examine the statistical properties of these regions, including the degree distribution. In the following sections, we begin by introducing the network construction method and the evaluated characteristics, and then describe how these magnetic patches are tracked.

\subsection{Identification of magnetic elements}\label{sec:identification}
For modeling the magnetic field structure in solar images, each pixel is considered as a node in the network. Then, edges are defined between these nodes using a modified version of the Visibility Graph. According to this method, two pixels are connected only if the magnetic field strength at both pixels exceeds the maximum absolute field strength along the direct path between them:

\begin{equation}
I_{i_1,j_1},\ I_{i_2,j_2} > \max_{\text{path}} |Ic|
\end{equation}
where $I_{i1,j1}$ and $I_{i2,j2}$ are the unsigned magnetic intensities (absolute values of BLOS) of any two arbitrary pixels with
different polarities, and $I_{c}$ corresponds to the maximum absolute value of all pixels placed along the line joining the two pixels.  This condition ensures that only connections are made between pixels with opposite polarities. 

To remove noise and prevent low-importance data from entering the network, pixels with a magnetic intensity lower than 12 Gauss are filtered out. The final network is defined as a directed, weighted graph, where the weight of the edges corresponds to the $B_{LOS}$ value at the source pixel.
\begin{figure*}
\centering
		\includegraphics[width=14cm,height=7cm]{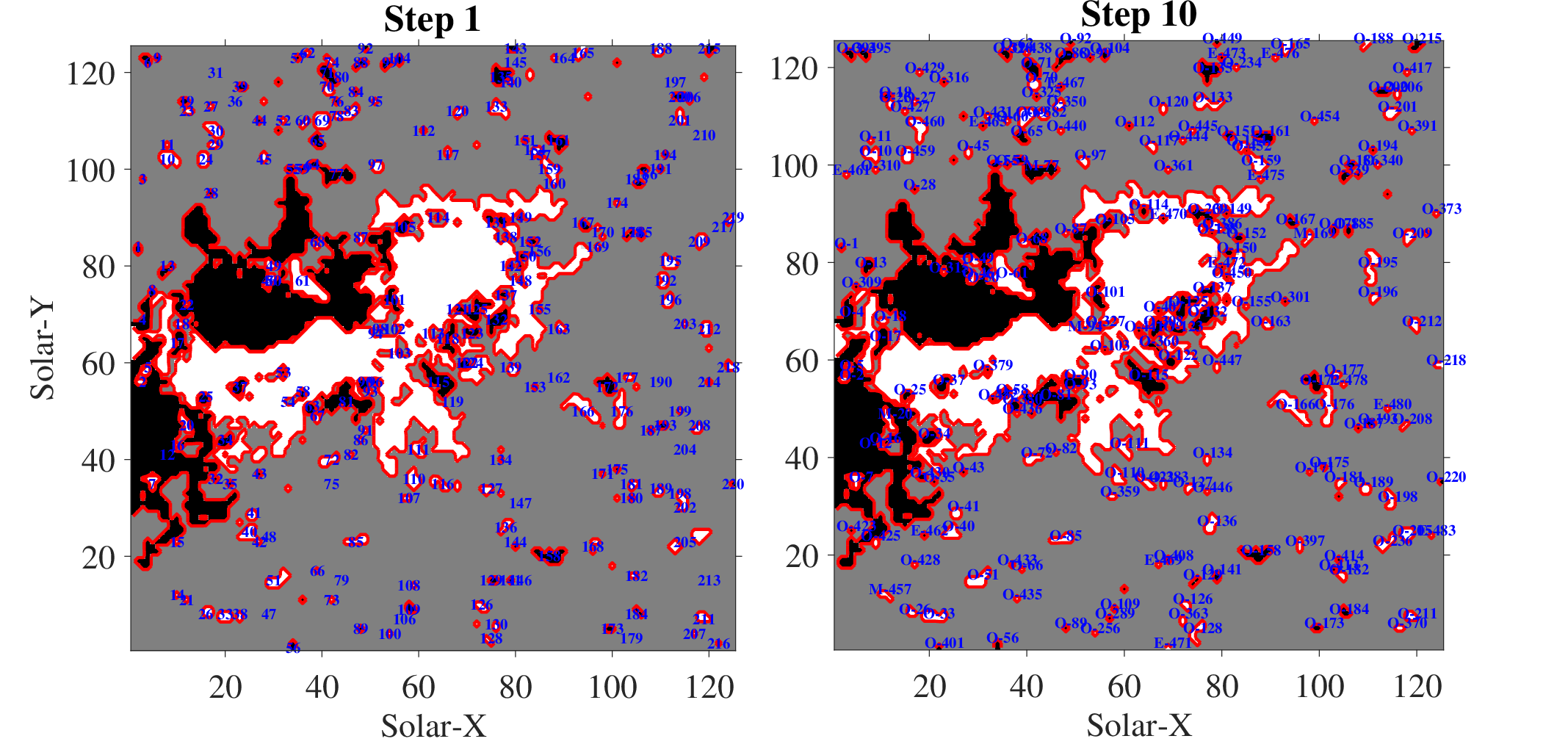}
		\caption{The left panel displays an HMI cutout image of AR NOAA No., 1158, captured at 16:59 on February 14, 2011. The right panel shows the corresponding magnetogram taken $7.5$ minutes later. Both images cover a field of view of $125 \times 125$ pixels. Magnetic patches are identified using a threshold of 18 Gauss, and a complex network is constructed accordingly. Patch boundaries are marked in red, while blue labels indicate their assigned IDs. Based on the implemented tracking algorithm, patches such as $O-2$, $O-4$, and $O-5$ are identified as one-to-one matches across the two frames. Some patches, like $77$ and $64$, undergo merging and are tracked as $M-77$. New patches (e.g., $E-462$ and $E-471$) appear, while others, such as 14 and 21, disappear in the later frame.
}
		\label{fig:track}
	\end{figure*} 
\subsection{Magnetic Element Tracking} \label{sec:tracking}

After identifying magnetic regions through the network-based method (Section~\ref{sec:identification}), a tracking procedure is employed to establish temporal correspondence between regions in consecutive magnetograms. This step aims to monitor the stability and evolution of magnetic structures over time.\\
The identification of temporal correspondence is based on comparing physical properties of magnetic regions in consecutive frames, including their centroid location, area, and polarity. The data consist of a series of line-of-sight magnetograms with a temporal cadence of 45 seconds.
The tracking algorithm relies on two primary criteria to match regions across time:\\
\begin{enumerate}	
	\item \textbf{Spatial overlap} - the two regions must share a sufficient number of common pixels, with the overlapping area exceeding a threshold (at least 30\% of their area);
	
	\item \textbf{Centroid proximity} - the distance $D$ between the flux-weighted centroids of the two regions must satisfy the following condition:
	\begin{equation}
		D = \sqrt{\max(A_l, A_m)}
		\label{track}
	\end{equation}
	where $A_l$ and $A_m$ denote the areas of the candidate regions at times $t_k$ and $t_{k+1}$, respectively. The centroid distance must be less than $D$ for the regions to be considered matched.	
\end{enumerate}

When both criteria are satisfied, the regions are considered temporally linked and assigned the same label. Otherwise, specific events such as appearance, disappearance, merging, or fragmentation are inferred based on spatial relations across frames. For instance, if multiple regions in a frame are found to correspond to a single region in the next frame, a merging event is identified; conversely, if one region splits into several, fragmentation is inferred. Emergence or disappearance is assumed when there is no spatial correspondence in the neighboring frames.

This approach allows for reconstructing the full temporal evolution of magnetic elements and provides a robust framework for analyzing the dynamics of the photospheric magnetic field. For a demonstration of this procedure, see the example illustrated in Figure \ref{fig:track}, which shows the evolution of region labels over a three-minute interval.

\begin{figure}[ht]
    \centering
    \includegraphics[width=1.05\textwidth]{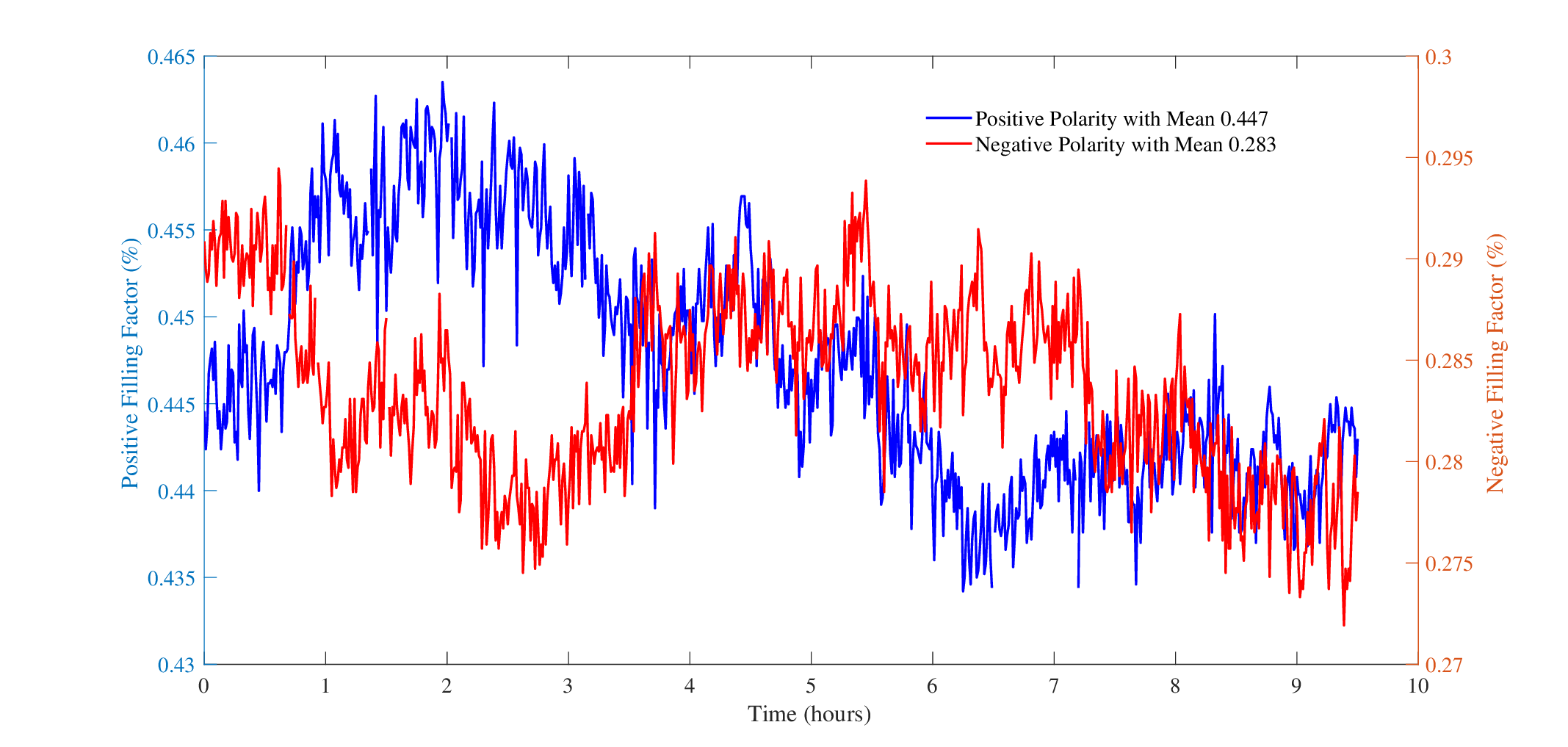} 
    \caption{ Filling factors of positive (red) and negative (blue) polarities of magnetic elements over 570.75 minutes. The Pearson and Spearman correlation coefficients are -0.101 and -0.090, respectively.}
    \label{fig:filing}
\end{figure}

\begin{figure}[htb]
  \centering
  \includegraphics[width=1.05\linewidth]{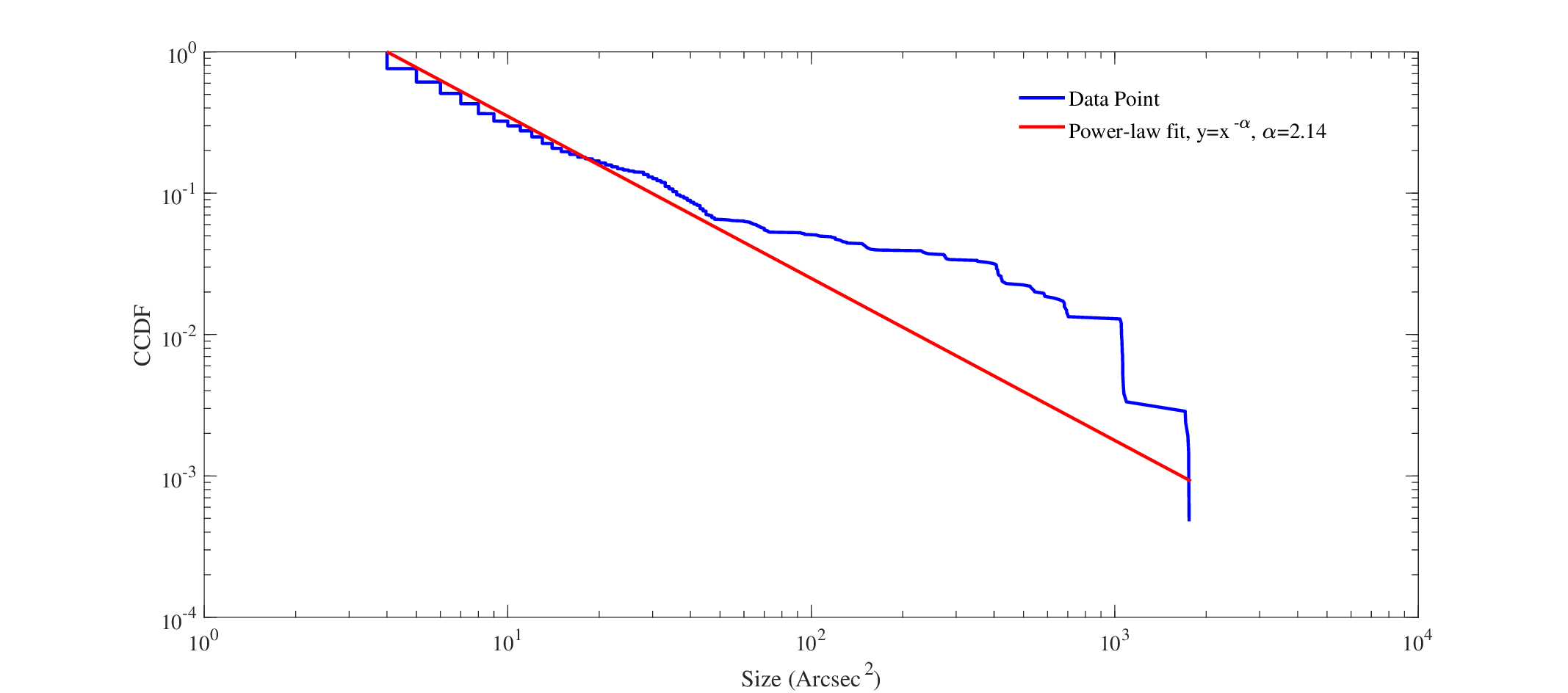}
  \caption{CCDF of the areas of magnetic patches in log-log scale. The red line shows a power-law fit with exponent $\alpha_{A}$ =2.14, using the MLE method. The blue curve represents the data points.}
  \label{fig:size_ccdf}
\end{figure}

\section{Results and Discussion}\label{result}
This section presents the results of applying the complex-network-based approach described in Section \ref{sec:method} to a 570.75 minutes sequence of HMI/LOS magnetograms of AR NOAA No., 1158,  with a cadence of 45 seconds. To reduce projection effects, a 125$\times$125 pixel subregion near the disk center was selected and processed. After applying a magnetic threshold of 18 Gauss, the method detected 16,313 independent magnetic patches of both polarities. The statistical analysis focuses on key physical properties of the patches (namely, area, magnetic flux, and lifetime) as well as their temporal evolution. These measurements provide important insight into the spatial complexity and dynamic behavior of magnetic structures in solar ARs, offering valuable input for understanding solar magnetic activity.
    
Figure~\ref{fig:filing} shows the temporal evolution of the filling factors for positive and negative polarity patches over a 570.75 minutes observation period. The plot uses a dual-axis format: the right vertical axis corresponds to the blue curve (positive polarity), while the left vertical axis corresponds to the red curve (negative polarity). The filling factor is defined as the fraction of the field-of-view (FOV) covered by magnetic patches of a given polarity, expressed as a percentage. Here, the FOV corresponds to a 125 $\times$ 125 pixel subregion. The time series reveals an average coverage of $44.7\%$ for positive patches and $28.3\%$ for negative ones. The Pearson and Spearman correlation coefficients between the two time series are $-0.101$ and $-0.090$, respectively, indicating a weak and statistically insignificant anti-correlation. This suggests that the two polarities evolve somewhat independently in the studied AR.

The size distribution of magnetic patches is an important diagnostic of underlying photospheric dynamics. Figure~\ref{fig:size_ccdf} presents the complementary cumulative distribution function (CCDF) of the patch areas on a log-log scale. A power-law fit, shown as a red line, is fitted to data points (blue curve) for values greater than $x_{\min} = 4~\mathrm{arcsec}^2$, yielding a scaling exponent of $\alpha_{A} = 2.14$ as estimated using the Maximum Likelihood Estimation (MLE) method. This power-law scaling indicates a scale-free organization of magnetic structures, suggesting that small-scale magnetic fields emerge and interact in a self-similar manner across a wide range of spatial scales. 

\begin{figure}[htb]
  \centering
  \includegraphics[width=1.05\linewidth]{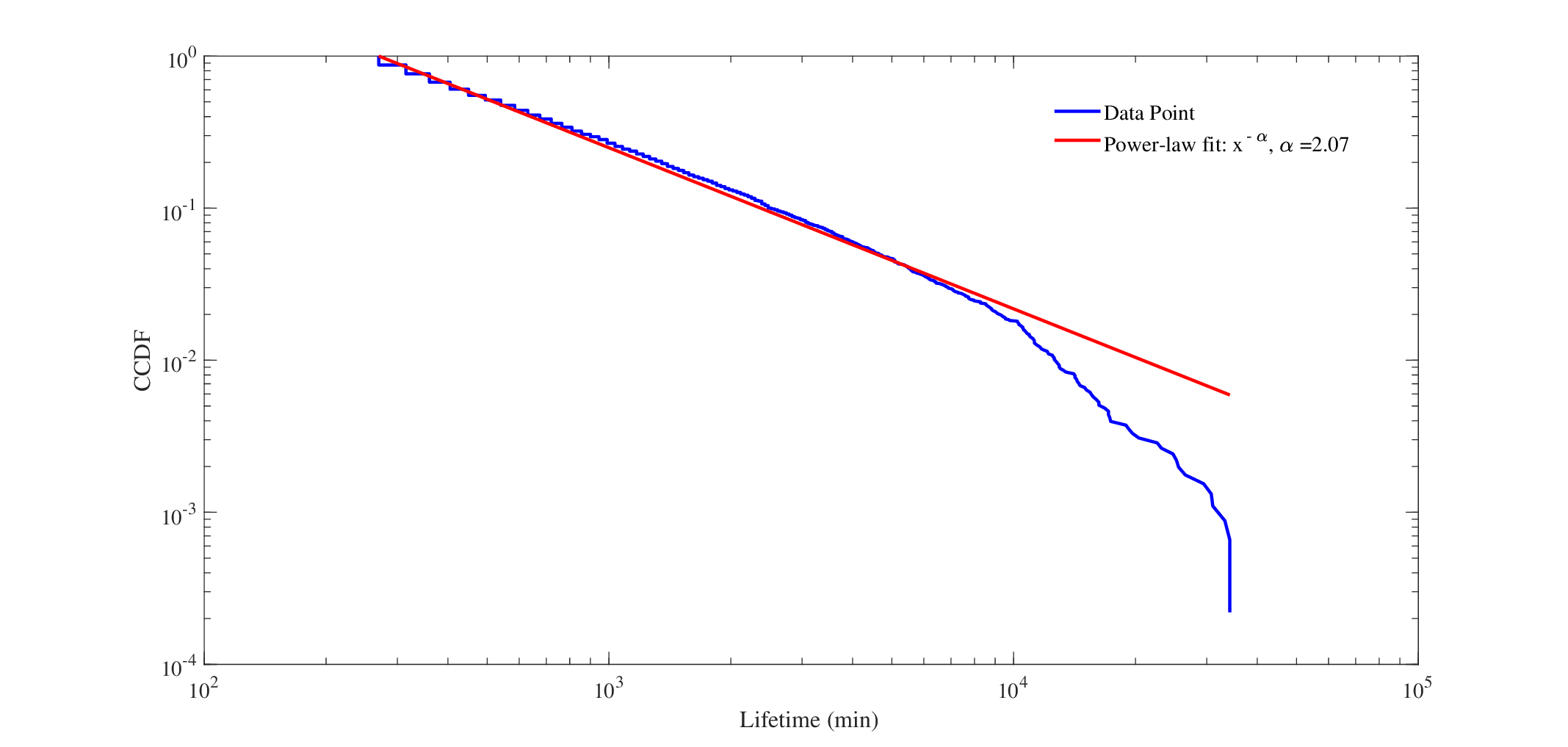}
  \caption{CCDF of patch lifetimes in a solar AR. The blue curve represents the empirical data points, and the red line shows a power-law fit with exponent $\alpha_{L} = 2.07$ using the MLE method, for lifetimes greater than $x > 270$ minutes. The heavy-tailed nature suggests the presence of long-lived patches.}
  \label{fig:lifetime}
\end{figure}

\begin{figure}[htb]
  \centering
  \includegraphics[width=1.05\linewidth]{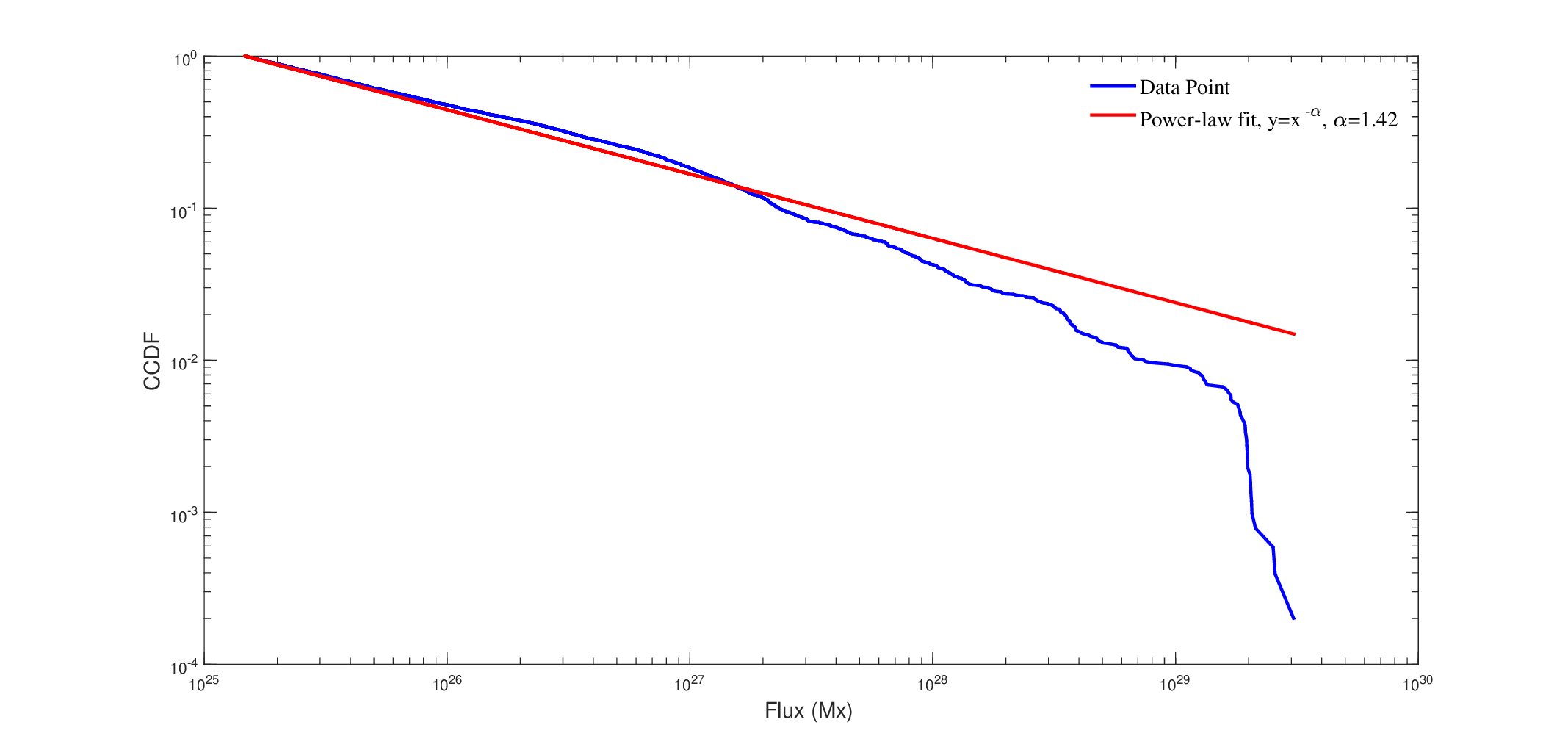}
  \caption{CCDF of magnetic flux for all patches in AR NOAA No., 1158. A power-law behavior with exponent $\alpha_{F} = 1.42$ is observed, indicating a scale-invariant distribution. The blue curve represents the empirical data points, while the red line shows the power-law fit.}
  \label{fig:flux_ccdf}
\end{figure}

\begin{figure}[ht]
\centering
\includegraphics[width=1.05\textwidth]{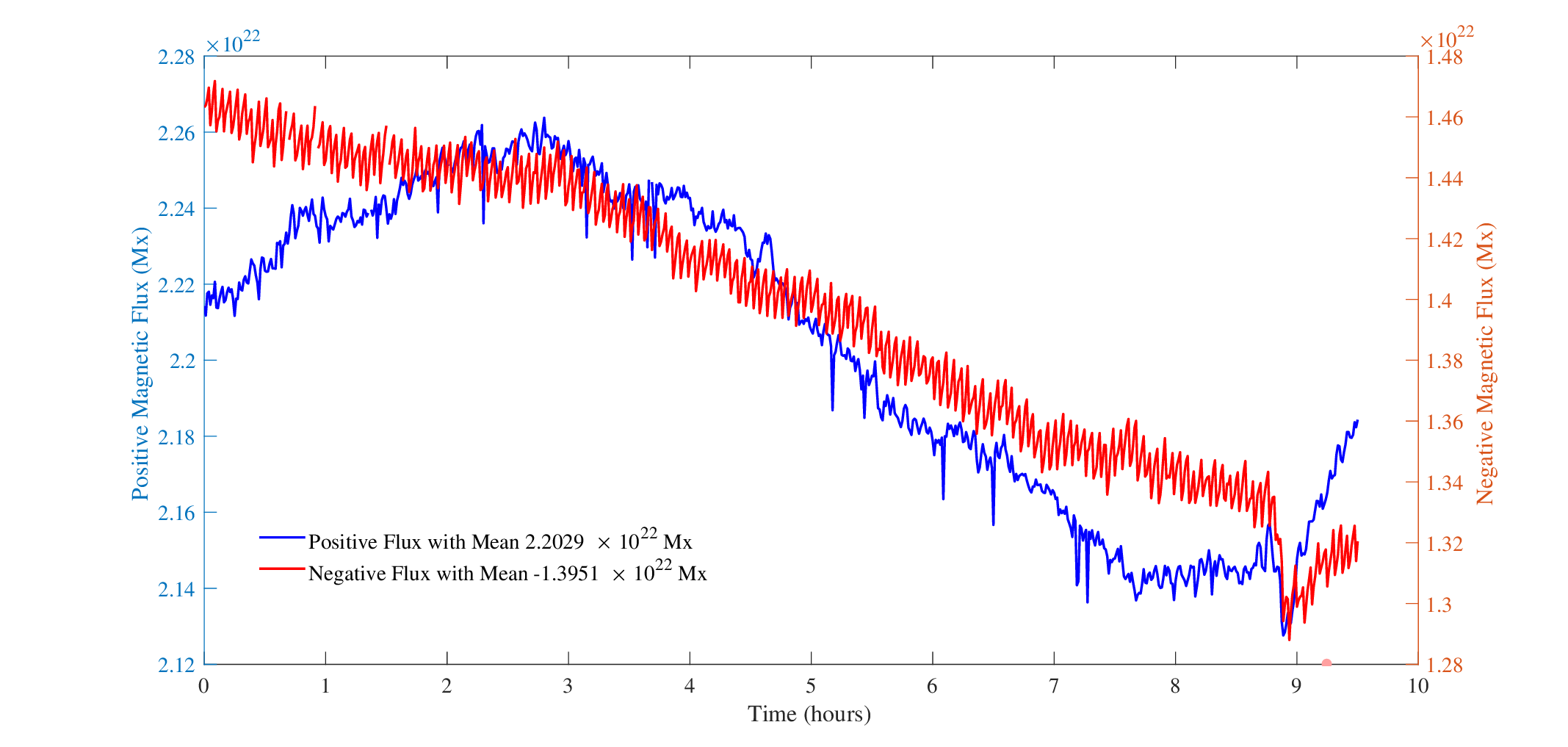}
\caption{Time series of total magnetic flux for positive (the red curve) and negative (the blue curve) polarities over the observation period. The plot uses a dual-axis format: the left vertical axis corresponds to the positive flux, while the right vertical axis corresponds to the negative flux. Pearson coefficient: $-0.57$, Spearman coefficient: $-0.82$, indicating a significant anti-correlation.}
\label{fig:flux-evolution}
\end{figure}

\begin{figure}[htb]
  \centering
  \includegraphics[width=1.05\linewidth]{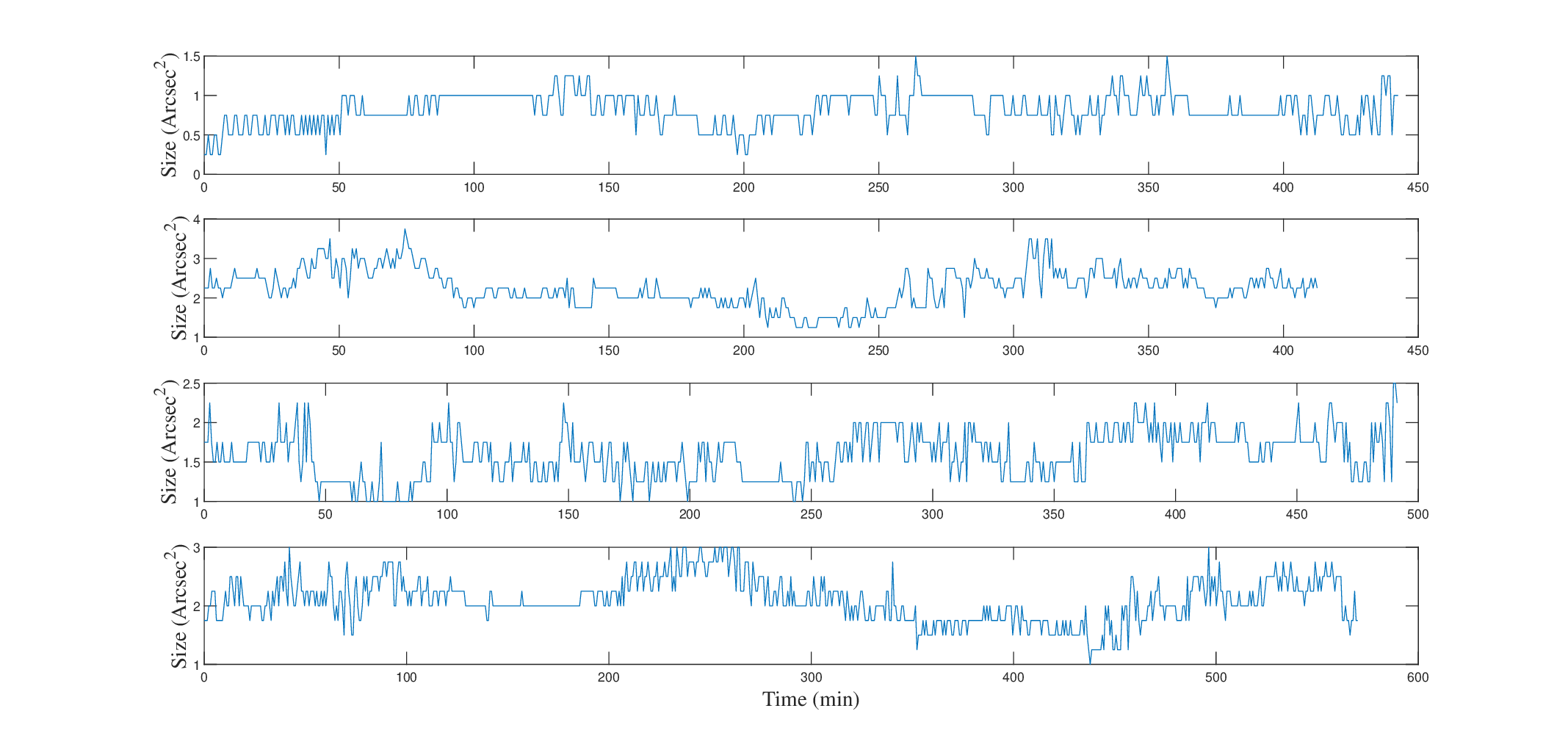}
  \caption{ Size variations of four magnetic patches with lifetimes longer than six hours, showing patterns of gradual and abrupt changes. These fluctuations are indicative of dynamic internal processes such as fragmentation, merging events, or localized flux interactions.}
  \label{fig:size_varition}
\end{figure}

\begin{figure}[htb]
  \centering
  \includegraphics[width=1.05\linewidth]{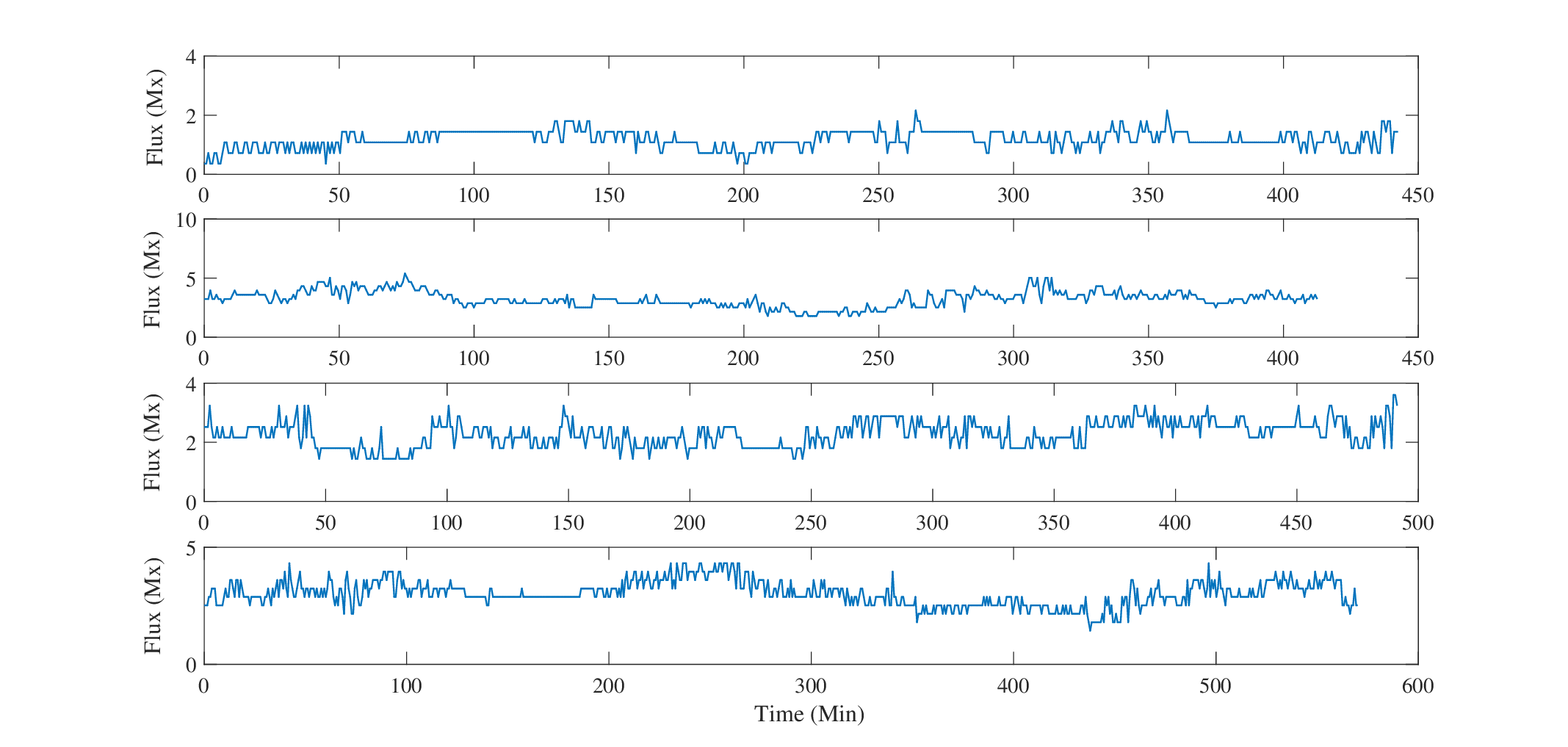}
  \caption{Flux variations of the same four magnetic patches shown in Figure~\ref{fig:size_varition}, highlighting a strong correlation with size changes. Abrupt jumps in flux may correspond to merging events or sudden flux emergence, while smooth variations could reflect slow diffusion or decay processes.}
  \label{fig:flux_varition}
\end{figure}

\begin{figure}[htb]
  \centering
  \includegraphics[width=1.05\linewidth]{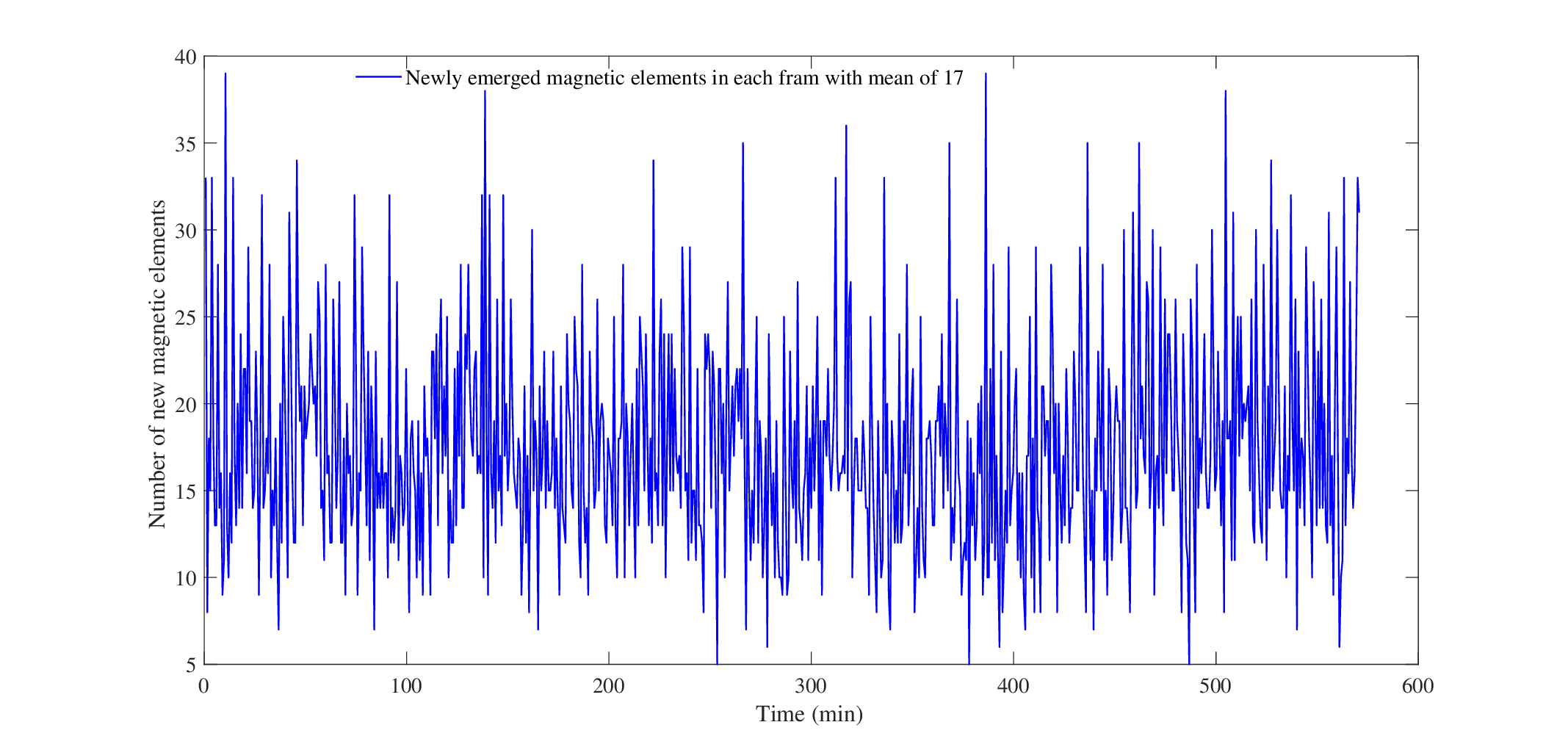}
  \caption{Number of newly emerged magnetic elements per frame during the 570.75 minutes. The Hurst exponent $H = 0.57$ suggests long-range persistence in the emergence pattern.}
  \label{fig:hurest}
\end{figure}

The slope obtained in this study ($\alpha_{A}=2.14$) is steeper than that reported by \citep{abramenko} ($\alpha \approx 1.9$), who analyzed ARs using traditional segmentation techniques. The larger exponent in our case may reflect the improved resolution and sensitivity of the network-based detection method, which can better isolate smaller-scale patches that might have been missed in earlier approaches.\\
Figure~\ref{fig:lifetime} displays the CCDF of the lifetimes of magnetic patches. The power-law fit was performed for lifetimes exceeding $x_{\min} = 270$ minutes, resulting in a scaling exponent $\alpha_{L} = 2.07$, estimated using the MLE method. The heavy-tailed nature of the resulting distribution suggests that, while most patches are short-lived, a non-negligible number can persist for several hours. This may reflect the stability of large-scale magnetic structures and their influence on the long-term organization of the AR.
The total magnetic flux carried by each patch provides valuable information about the magnetic energy distribution within the AR. Figure~\ref{fig:flux_ccdf} presents the CCDF of the total magnetic flux per patch in log-log scale. The distribution exhibits a clear power-law behavior for values above $x_{\min} = 1.47 \times 10^{25}~\mathrm{Mx}$, with a fitted exponent of $\alpha_{F} = 1.42$ obtained via the MLE method. The scale-invariant nature of the flux distribution—commonly observed in solar magnetic fields—indicates that only a small fraction of patches with high flux levels dominate the overall magnetic energy budget. Such distributions are characteristic of systems governed by self-organized criticality (SOC), where no preferred scale exists in the emergence processes. This implies that the flux distribution is shaped by dynamic interactions across multiple spatial scales, leading to a robust and scale-free organization of magnetic structures within the AR.     
Interestingly, this exponent is smaller than the value $\alpha \approx 1.85$ reported by \citep{parnell2009power}, who analyzed a large population of magnetic patches across the entire solar disk, including both active and quiet-Sun regions. The flatter slope found in our study indicates that in an AR like NOAA NO., 1158, strong-flux patches are relatively more common, contributing to a heavier-tailed distribution. Furthermore, \citep{abramenko} qualitatively showed that ARs with flatter magnetic power spectra tend to be more flare-productive. Although their study did not directly compute flux distributions, our results align with their findings by revealing the dominance of high-flux, complex structures in magnetically active environments.
The self-similar or self-organized criticality (SOC) behavior observed in magnetic patches can be regarded as a fundamental feature of the Sun's complex magnetic system during dynamic processes. This behavior underlies a wide range of solar atmospheric phenomena—from large-scale flaring activity to the smallest-scale brightenings such as nanoflares \citep{Farhang2018,kaki2022evidence}, microflares, campfires \citep{alipour2022automatic}, and bright points \citep{alipour2015statistical, rad2021energetics}. Notably, these events exhibit power-law distributions in their size and energy release, a hallmark of SOC systems \citep{lu1991avalanches,aschwanden2011self,georolis}. Such scale-invariant characteristics suggest that the solar corona operates near a critical state, where small magnetic disturbances may trigger energy releases over multiple scales \citep{bak1987self,charbonneau2001avalanche}. Therefore, SOC behavior in magnetic patches offers a powerful framework to understand solar atmospheric activity's statistical properties and scale-free nature \citep{tajik2023behavior}.\\
Figure \ref{fig:flux-evolution} shows the total magnetic flux of positive (the blue curve) and negative (the red curve) polarities over time. The mean values are $2.20 \times 10^{22}$ Mx for positive polarity and $-1.40 \times 10^{22}$ Mx for negative polarity, indicating a net positive flux in the AR. The Pearson and Spearman correlation coefficients are $-0.57$ and $-0.82$, respectively, suggesting a moderately strong anti-correlation, particularly nonlinear in nature. This flux imbalance and coupling may result from asymmetric flux emergence, cancellation, or redistribution.
To investigate the evolution of individual magnetic elements, Figure \ref{fig:size_varition} illustrates the size time-series for four patches with lifetimes exceeding six hours. The plots reveal both gradual and abrupt size changes. Such fluctuations are likely associated with internal processes like fragmentation, merging, or local flux interactions, reflecting the dynamic nature of magnetic field evolution in ARs.
Figure~\ref{fig:flux_varition} displays the flux evolution of the same four long-lived patches. The flux curves closely match the size changes shown in Figure~\ref{fig:size_varition}. The flux curves exhibit close correlation with the corresponding size changes shown in Figure \ref{fig:size_varition}. Abrupt jumps may correspond to merging events or sudden flux emergence, while smooth variations could reflect slow diffusion or decay processes. The temporal coherence between flux and area suggests a physical link between magnetic field strength and patch structure.
Figure~\ref{fig:hurest} shows the time series of newly emerged magnetic elements per magnetogram frame over 570.75 minutes, with an average emergence rate of 17 patches per frame. To assess the temporal structure of this emergence behavior, a Detrended Fluctuation Analysis (DFA) was performed, yielding a Hurst exponent of $H = 0.57$. This value indicates the presence of long-range temporal correlations, meaning that periods of high activity tend to cluster together. Such statistical persistence points to a memory effect in the underlying magneto-convective processes, suggesting that new emergence events are not entirely random but are influenced by previous magnetic dynamics.

\section{Conclusion} \label{sec:conclusion}
In this study, we applied a complex-network-based method for the identification and tracking of magnetic patches on the solar photosphere. Using a 570.75 minutes time series of high-resolution LOS magnetograms from HMI/SDO with a cadence of 45 seconds, we analyzed the evolution of magnetic structures in AR NOAA No., 1158. \\
The statistical analysis revealed that patch areas, lifetimes, and total unsigned magnetic fluxes follow power-law distributions above clear thresholds. Specifically, the power-law exponent was found to be $-2.14$ for patch area, $-2.07$ for lifetime, and $-1.42$ for magnetic flux. These heavy-tailed distributions confirm that magnetic structures in the AR are organized in a scale-free manner, consistent with expectations from self-organized criticality and turbulent convection models.\\
By constructing time series of total flux and filling factor, we observed that while the AR exhibited an overall trend of increasing activity, it also displayed substantial temporal fluctuations, indicative of intermittent processes. The emergence rate of new magnetic patches was quantified and shown to have long-range temporal correlations, as captured by a Hurst exponent of $H = 0.57$. This value indicates a persistent memory effect in the patch emergence process, potentially linked to underlying convective motions in the sub-photospheric layers.\\
We also tracked the evolution of long-lived patches and identified both stable and unstable behavior. Some patches exhibited gradual growth and decay, while others underwent rapid fragmentation or merging events. These behaviors reflect the underlying complexity of magnetic interactions in active regions and support the view that magnetic field evolution is governed by nonlinear, scale-invariant processes, potentially driven by self, organized criticality and turbulent magneto-convection. Future studies could extend this approach to longer time series and apply it to quiet-Sun regions or multiple active regions, enabling comparative analyses that may uncover universal or environment-dependent mechanisms shaping the solar photospheric magnetic field.

\section*{Acknowledgment}
The authors would like to thank the NASA/SDO science group for making data publicly available. 


\bibliographystyle{aasjournal}
\bibliography{ref_Tajik.bib}

\end{document}